%% file: main.tex
\documentclass[floatfix,amsmath,superscriptaddress,amssymb,prb]{revtex4-2}

\usepackage{graphicx}
\usepackage{dcolumn}
\usepackage{bm}
\usepackage{physics}
\usepackage{hyperref}
\hypersetup{colorlinks,citecolor=blue,linkcolor=blue,urlcolor=blue}
\usepackage{xcolor}
\usepackage{ulem} 
\usepackage{dsfont}
\usepackage{xr}
\usepackage{cleveref}
\usepackage{comment}

\usepackage{caption}
\usepackage{subcaption}

\newcommand{\kp}{$\mathbf{k}\cdot\mathbf{p}$ }

\newcommand{\DG}{\Delta_\Gamma}
\newcommand{\Dh}{\Delta_\mathrm{hyb}}
\newcommand{\HBHZ}{H_\text{\tiny BHZ}}
\newcommand{\nwkpy}{\textsf{nwkp$_\textsf{y}$}}  
\newcommand{\rprp}{\boldsymbol{r}_{\perp}}

\begin{document}

\title[A semimetal state in nanowires]
  {Spin-orbit control of Dirac points and end states in inverted gap nanowires}

\author{Andrea Vezzosi}
\email{andrea.vezzosi@unimore.it}
\affiliation{Dipartimento di Fisica, Informatica e Matematica, Universit\`a di Modena e Reggio Emilia, Via Campi 213/a, 41125 Modena, Italy}
\affiliation{Istituto Nanoscienze CNR, Via Campi 213/a, 41125 Modena, Italy}
\author{Andrea Bertoni}
\email{andrea.bertoni@nano.cnr.it}
\affiliation{Istituto Nanoscienze CNR, Via Campi 213/a, 41125 Modena, Italy}
\author{Marco Gibertini}
\email{marco.gibertini@unimore.it}
\affiliation{Dipartimento di Fisica, Informatica e Matematica, Universit\`a di Modena e Reggio Emilia, Via Campi 213/a, 41125 Modena, Italy}
\affiliation{Istituto Nanoscienze CNR, Via Campi 213/a, 41125 Modena, Italy}
\author{Guido Goldoni}
\email{guido.goldoni@unimore.it}
\affiliation{Dipartimento di Fisica, Informatica e Matematica, Universit\`a di Modena e Reggio Emilia, Via Campi 213/a, 41125 Modena, Italy}
\affiliation{Istituto Nanoscienze CNR, Via Campi 213/a, 41125 Modena, Italy}
\date{\today}
 
\setlength{\parskip}{0pt} 

 \begin{abstract}
 We predict that in InAs/GaSb nanowires with an inverted band alignment a transverse electric field induces a collapse of the hybridization gap, and a semimetal phase occurs. We use a self-consistent \kp approach and an adapted Bernevig-Hughes-Zhang model to show that massless Dirac points result from exact cancellation between the kinetic electron-hole coupling and the field-controlled spin-orbit coupling. End states -- mid-gap states localized at the extremes of a finite nanowire -- are supported up to a critical field, but suddenly fade away as the system is driven through the semimetal phase, eventually evolving to trivial surface states, which expose a spin-orbit induced topological transition to the normal phase.
\end{abstract}

\maketitle

InAs/GaSb heterointerfaces feature a broken gap alignment, with the bulk conduction band (CB) of InAs lying below the bulk valence band (VB) of GaSb \cite{vurgaftman2001}. In their two-dimensional (2D) form, such as a InAs/GaSb bilayer embedded in a insulating material\cite{LiuPRL08}, layer widths can be engineered such that confinement energies partially compensates for the large ($\sim$$0.15\,\text{eV}$) inverted alignment. If only a few valence subbands are tuned above the lowest conduction subband, an overall hybridization gap opens at the crossings between subbands with opposite curvature.

Most of the interest in inverted InAs/GaSb quantum wells (QWs) stems from being prototypical 2D topological insulators (TIs) \cite{Konig2008,Hasan2010a,Maciejko2011,Ando2013,Hsu2021} 
due to the  non-trivial topology of the occupied Hilbert space of the inverted insulating state. In this regime, bulk-boundary correspondence leads to a semimetal occupied state within the bulk gap, in the form of a massless Dirac cone at the $\Gamma$ point, due to conductive edge states travelling around the boundary of a finite sample, which are spin-momentum locked and protected from scattering by the bulk gap if time-reversal (TR) symmetry is preserved. This phenomenon, known as the Quantum Spin Hall (QSH) Effect \cite{Murakami2004,Maciejko2011}, holds potential for groundbreaking applications in electronics and quantum technologies  \cite{Tokura2017,Han2018}. An important aspect is that, since electrons and holes are hosted in separate layers, the band inversion can be switched to normal ordering by a vertical electric field, driving the system across the topological phase transition \cite{LiuPRL08, KnezPRL11, Qu2015}.

Investigations have been extended to one-dimensional (1D) systems \cite{Vinas2017,KishorePRB12,LuoSciRep16} grown in the form of core-shell InAs/GaSb nanowires (NWs) \cite{Clerico2024,Wang2024Vac}, showing that inverted gap NWs support an overall hybridization gap (in the order of a few meV) for sufficiently thin core and shell \cite{LuoSciRep16}. Unlike their 2D counterparts, InAs/GaSb NWs are not supposed to be genuine TIs due to the lack of particle-hole (PH) or charge-conjugation symmetry that are needed to protect proper topological states in 1D~\cite{Hasan2010a}. However, mid-gap zero-dimensional states, termed end states, are predicted to be localized at the extremes of a finite band-inverted NW \cite{VinasPRB20}, as remnants of a topological invariant which is expected for 1D systems featuring exact PH symmetry \cite{Hasan2010a}. 

Since GaSb is a strong spin-orbit (SO) coupled material, large spin splittings are induced by a transverse electric field which breaks spatial inversion symmetry. Theoretical \cite{ZakharovaJETP2011} and experimental \cite{NichelePRL17} studies of the resulting spin-dependent hybridization gap have indeed been performed in InAs/GaSb QWs and full SO polarization has been achieved. However, no sign of gap collapse induced by the field via SO coupling, which is the focus of the present paper, has been reported. Moreover, SO effects in inverted gap NWs have not been investigated so far.

In this Letter we predict that in band inverted InAs/GaSb NWs a transverse field induces a semimetallic phase, as a result of the collapse of the hybridization gap between oppositely spin-polarized subbands. This is traced to SO coupling, via removal of inversion symmetry operated by the field. Differently from the effect of the field in 2D QWs~\cite{LiuPRL08}, which is electrostatic in nature, in NWs the field couples with SO which, at a critical value, exactly compensates the VB-CB kinetic coupling, making the gap to vanish.
We show further that localized end states are supported in the low-field regime up to the critical field, but suddenly disappear, eventually evolving into trivial surface states merging with the CB and VB continua. Remarkably this suggests the occurrence of a 1D, SO-induced topological transition, which is preserved despite the effective breaking of the underlying protecting symmetry.

\begin{figure}[ht]
    \centering
    \includegraphics[width=1\columnwidth]{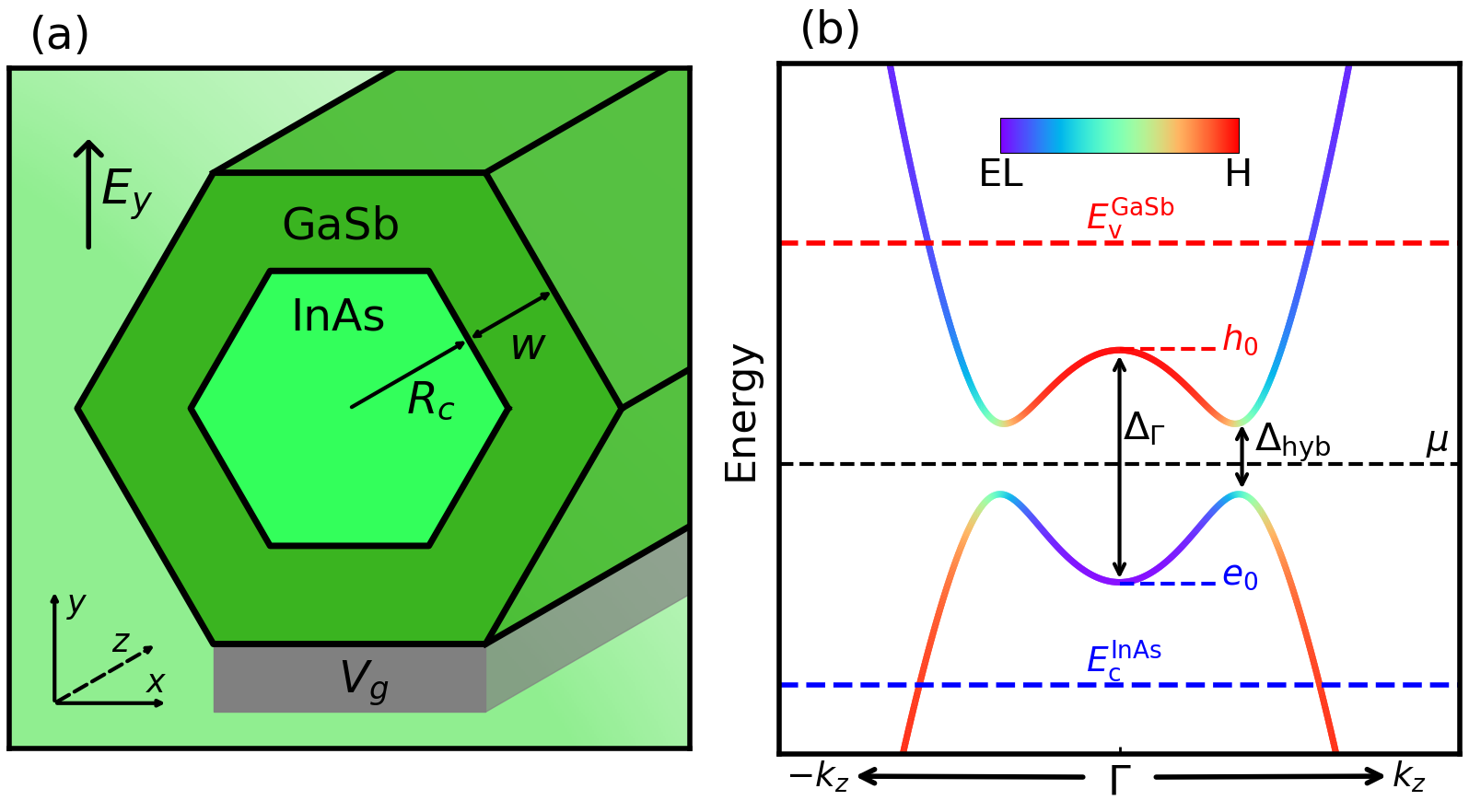}
    \caption{(a) Sketch of a InAs core, with edge-to-edge radius $R_c$ covered by a GaSb shell of width $w$ and an attached bottom gate. (b) Schematic band structure in a weakly inverted phase. Blu and red dashed lines indicate the VB ($E_v^\mathrm{GaSb}$) and CB ($E_c^\mathrm{InAs}$) inverted bulk band edges, a black dashed line indicate the chemical potential $\mu$. Labels indicate the inverted electron ($e_0$) and hole ($h_0$) lowest subbands, and the inverted ($\DG$) and the hybridization ($\Dh$) gaps, see text. The hue of the subbands represents the character, from pure $s$ (EL, blue) to pure $p$ (H, red) character.}
    \label{fig:fig1}
\end{figure}

Our prototype heterostructure is shown in Fig.~\ref{fig:fig1}(a). An InAs [111]-grown core with radius $R_c$ is embedded in a thin GaSb shell of width $w$. We choose the values of $R_c$ and $w$ so that quantum confinement almost compensates the inverted alignment of the bulk CB/VB edges $E_{\rm c}^{\rm InAs}/E_{\rm v}^{\rm GaSb}$, and only one spin pair of conduction and valence subbands, with edges $e_0$ and $h_0$, respectively, are inverted, as sketched in Fig.~\ref{fig:fig1}(b). Due to kinetic CB-VB mixing, as determined by the optical matrix element $P$, the two subbands anticross, and a hybridization gap $\Dh$ opens at two Kramers-related finite in-wire wave vectors. $\Dh$ is typically smaller then the inverted gap at the $\Gamma$ point, $\DG = \left|e_0 - h_0\right|$, and the material is in an insulating phase \cite{LuoSciRep16}. Note that in inverted heterostructures, electrons may lower their energies by moving from the higher VB to the lower CB until the chemical potential $\mu$ balances across the layers; for an isolated system $\mu$ is determined by the charge neutrality condition, otherwise $\mu$ is fixed by contacts. In either case, the resulting band bending at the interfaces competes with quantum confinement and should be included for a quantitative evaluation of the gaps.

We perform \kp calculations with the 8-band Kane Hamiltonian \cite{Vezzosi2022,Clerico2024} with $R_c = 7\,\mathrm{nm}$ and $w = 4.88\,\mathrm{nm}$. The NW is invariant for translations along the growth direction $z$ and a constant electric field $E_y$ is applied along the normal direction $y$. The Hartree potential due to free electron and hole charge densities is included by the self-consistent solution of the Poisson equation, using Neumann boundary conditions at all facets but the bottom one, where Dirichlet boundary conditions are enforced which is meant to simulate the presence of a back-gate, as in Fig.~\ref{fig:fig1}(a). Calculations are performed using the \nwkpy\ library \cite{Vezzosi2022}. Bulk band parameters and details on the numerical calculations are given in Sec.~\ref{sec:kp calculations} of the Supporting Information (SI).

\begin{figure}[ht]
    \centering
    \includegraphics[width=1\columnwidth]{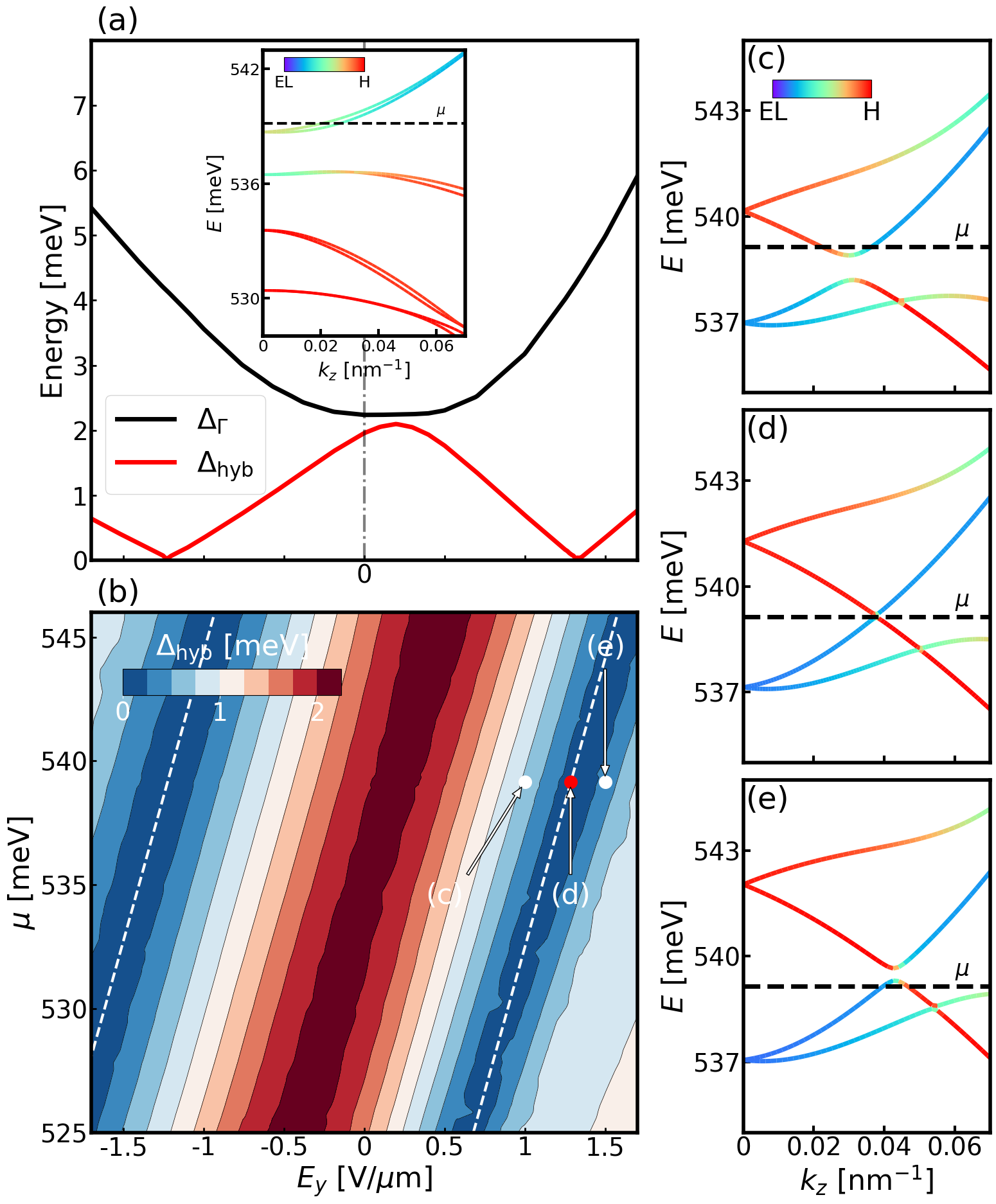}
    \caption{(a) Inverted gap $\Delta_{\Gamma}$ (black) and hybridization gap $\Dh$ (red) for a chemical potential $\mu$=539.14 meV as a function of the transverse electric field $E_{y}$. Inset: subband dispersion at $E_y=0$; hue is calculated from the spinorial solution of the \kp Hamiltonian according to Eqs.~(\ref{eq:electron_character}), (\ref{eq:hole_character}). (b) Hybridization gap $\Dh$ as a function of $\mu$ and $E_{y}$. White dashed lines highlight the points of vanishing gap. (c-e) Energy dispersion at the electric fields $E_y = 1, 1.285, 1.5$ V$/\mu m $,  and $\mu = 539.14$ meV, corresponding to the points in panel (b). $\mu$ has been chosen so that it lies at the Dirac point at the critical field $\overline{E}_+$}
    \label{fig:fig2}
\end{figure}

In the inset of Fig.~\ref{fig:fig2}(a) we show the band structure with no applied electric field, $E_y=0$. A small hybridization gap $\Dh\sim 2\,\mathrm{meV}$ opens between two slightly dispersive subbands, hence $\DG\simeq\Dh$. The hue of the bands shows that the subbands have a substantially hybridized character between EL ($s$-like) and H ($p$-like) character, from the $\Gamma$ point up to the anti-crossing, beyond which they attain a marked EL or H character.\footnote{The CB never attains a full EL character, as shown also in Ref.~\cite{LuoSciRep16}} Although $E_y=0$, each subband is slightly spin split away from $\Gamma$, since exact inversion symmetry is removed by the self-consistent field via the anisotropic boundary conditions imposed by the gate.

A finite electric field $E_y$ has several effects: \textit{i)} $e_0$ and $h_o$ move apart, increasing the inverted gap $\DG$, see black line in Fig.~\ref{fig:fig2}(a). Accordingly, the subbands become less hybridized and more dispersive around $\Gamma$; \textit{ii)} each subband undergoes strong spin splitting due to inversion symmetry removal brought about by $E_y$; \textit{iii)} the combined effect of \textit{i)} and \textit{ii)} lowers the hybridization gap $\Dh$, see red line in Fig.~$\ref{fig:fig2}(a)$. Surprisingly, despite the active electron-hole kinetic coupling, \textit{the gap precisely vanishes at two critical fields $\overline{E}_\pm \sim \pm 1.3\, V/\mu m$}, and opens for again for $|E_{y}| > |\overline{E}_{\pm}|$.

These trends are exemplified in Figs.~\ref{fig:fig2}(c-d), where we show the band structure at selected values of $E_y$, from slightly below to slightly above the positive critical field $\overline{E}_+$ and fixed $\mu$. For $E_y\lesssim \overline{E}_+$ [Fig.~\ref{fig:fig2}(c)] each subband is strongly spin split due to SOC, and one can identify two anticrossings between "external" subbands and between "internal" subbands, with substantially different anticrossing behaviours. At $E_y = \overline{E}_+$ [Fig.~\ref{fig:fig2}(d)] $\Dh$ vanishes in an otherwise overall gapped region. Hence the system is in a \textit{semimetallic} phase, with two subbands crossing at a finite wave vector $k_0$ with massless Dirac dispersion near the degeneracy point. Consistently with the vanishing hybridization gap, the crossing subbands show a pure EL and H character. Finally, if $E_y > E_+$ [Fig.~\ref{fig:fig2}(e)], the energy gap reopens. Clearly, due to Kramers degeneracy, a partner Dirac point exists at $-k_0$ with opposite spin orientations (not shown here).

The critical fields depend on the chemical potential $\mu$ through the self-consistent Hartree potential generated by the free charge distributions.
The overall behaviour of $\Dh$ is shown in Fig.~\ref{fig:fig2}(b) as a function of $E_y$ and $\mu$, showing that a vanishing $\Dh$ occurs at any $\mu$ in the considered range. Note a slight $E_+/E_-$ asymmetry [see also Fig.~\ref{fig:fig2}(a)] at fixed $\mu$, due to the presence of the bottom gate that removes exact inversion symmetry in the field direction.

The intimate reason underneath the collapse of the hybridization gap at $\overline{E}_\pm$ is a compensation of SOC with the kinetic CB-VB interaction. To show this we develop an analytical model that describes the states near the closing gap. We start from the Bernevig-Hughes-Zhang (BHZ) Hamiltonian \cite{BHZ2006} which describes the low-energy states (with respect to $\mu$) of a \textit{symmetric} QW \cite{Krishtopenko2019} (similarly to our centro-symmetric NW) in the minimal basis set of the spin-degenerate states of the QW at $\Gamma$, $\left\{\ket{\text{EL}+}, \ket{\text{H}+}, \ket{\text{EL}-}, \ket{\text{H}-}\right\}$. The BHZ Hamiltonian, extended to include a vertical electric field perpendicular to the QW \cite{rothe2010} which breaks the symmetry, reads
\begin{equation}
\HBHZ = \varepsilon(k) \mathds{1}_4 + 
    \begin{pmatrix}
         \mathcal{M}(k)&\mathcal{A} k_{+}& 0 & -S_{0} k_{-}^2\\
         \mathcal{A} k_{-}&- \mathcal{M}(k)&  S_0 k_{-}^2 & 0\\
         0 &  S_{0} k_{+}^2 & \mathcal{M}(k)&-\mathcal{A}k_{-} \\
         -S_0 k_{+}^2 & 0 & -\mathcal{A}k_{+} & -\mathcal{M}(k)
    \end{pmatrix} \,,
    \label{eq:BHZ_hamiltonian_2d}
\end{equation}
where $\mathds{1}_4$ is a $4 \times 4$ identity matrix, $k=(k_x,k_y)$ is the  wave vector in the QW plane, $k_{\pm} = k_x \pm i k_y$, $k^2=k_x^2 + k_y^2$, $\mathcal{M}(k) = M - B k^2$ and $\varepsilon(k) = C - D k^2$. The field-controlled Rashba term $S_{0}=\kappa\times e E_{z}$ removes structural inversion symmetry and couples EL and H states with opposite spin. $B, C, D, M$ and $\kappa$ are material and structure dependent parameters (see Tab.~\ref{tab:param_BHZ} of the SI).

\begin{figure}[ht]
    \centering
    \includegraphics[width=1\columnwidth]{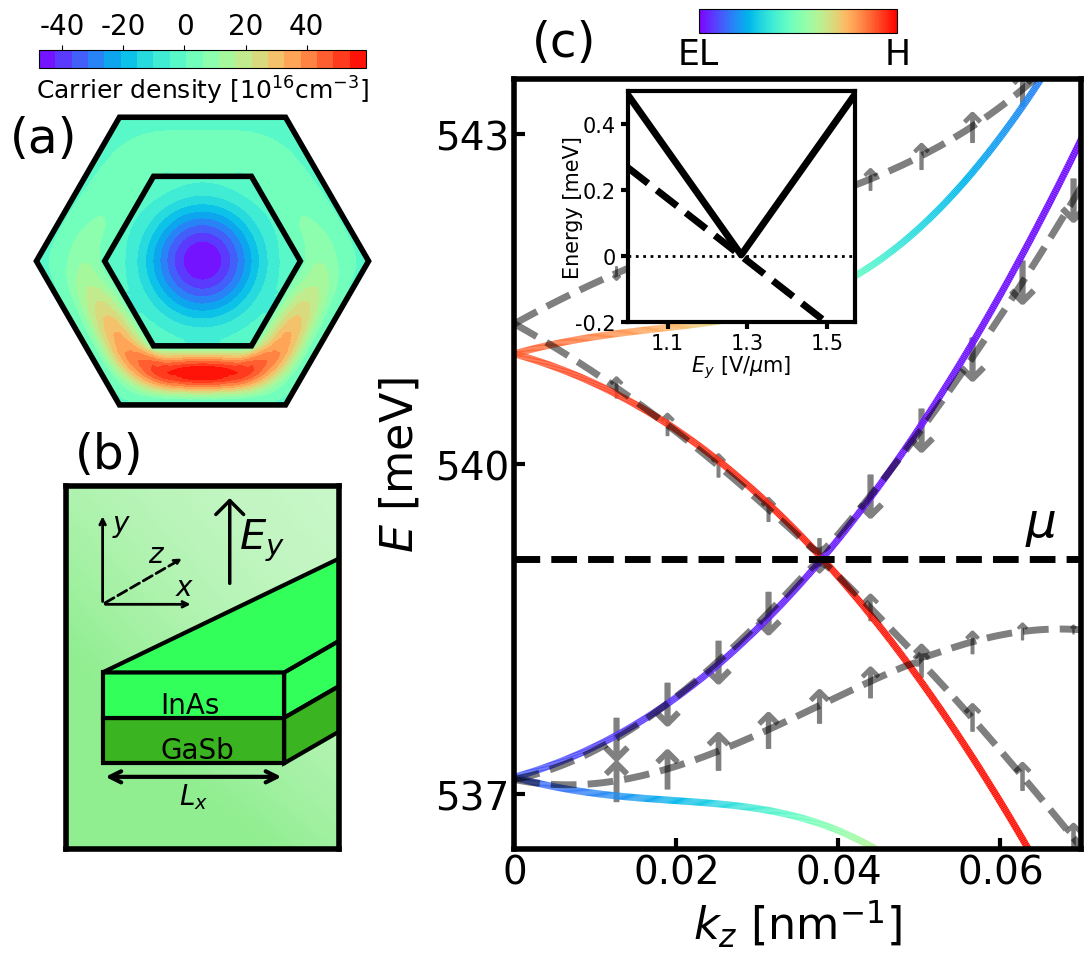}
    \caption{(a) Charge distribution at the critical field $\overline{E}_+=1.285$ V/$\mu$m and $\mu=539.144$ $e$V. (b) Sketch of a laterally confined GaSb/InAs QW 
    described by Eq.~(\ref{eq:BHZ_projected_n=1}). (c) Energy dispersion at the critical field $\overline{E}_y$. Full, colored lines: from Eq.~\ref{eq:BHZ_projected_n=1}. Dashed, gray lines: 8-band \kp  self-consistent calculation (reported from Fig.~\ref{fig:fig2}(d)), with the modulus of the $x$-component of the vector spin operator $\hat{\boldsymbol{S}}$, calculated as $S_x = \matrixel{\Psi}{\hat{S}_x}{\Psi}$ represented by the size of the upward (positive $S_x$) or downward (negative $S_x$) pointing arrows. Inset: Hybridization gap $\Dh$ (full line) and gap closing condition, LHS of Eq.~(\ref{eq:symmetry_conditions})(b) (dashed line), of the laterally confined BHZ model $vs$ $E_y$.}
    \label{fig:fig3}
\end{figure}

We now observe that at $\overline{E}_+$ holes in the GaSb layer are attracted near the bottom gate [see Fig.~\ref{fig:fig3}(a)] and the electronic system can be approximately mapped to a laterally confined GaSb/InAs QW [Fig.~\ref{fig:fig3}(b)]. Hence, we use $\HBHZ$ adding a lateral confinement in one direction. Rewriting Eq.~(\ref{eq:BHZ_hamiltonian_2d}) in the reference frame of Fig.~\ref{fig:fig1}(a), $(k_x,k_y)\rightarrow (k_z, -i \pdv{}{x})$, we project $\HBHZ$ onto the real space basis 
$
    \psi_n(x) = \sqrt{{2}/{L_x}} \sin\left({n \pi }/{L_x} \left(x-{L_x}/{2}\right)\right).
$
As we aim at describing the two low-energy doublets in Figs.~\ref{fig:fig2}(c-e), we limit ourself to a minimal basis with $n=1$, obtaining the $4 \times 4$ Hamiltonian
\begin{equation}
\begin{split}
   H_{4}(k_z) &= \varepsilon'(k_z) \mathds{1}_4 + \mathcal{M}'(k_z) \sigma_{0} \tau_z \\&+ \mathcal{A} k_z \sigma_z \tau_x + S_0 \left( k_z^2 - \frac{\pi^2}{L_x^2}\right) \sigma_y \tau_y 
\end{split} \,,
\label{eq:BHZ_projected_n=1}
\end{equation}
where $\varepsilon'(k_z) = C - D \left[ (\pi/L_x)^2 + k_z^2\right],
\mathcal{M}'(k_z) = M - B \left[ (\pi/L_x)^2 + k_z^2\right]$,
and $\tau_{x,y,z}$ and $\sigma_{0,x,y,z}$ are Pauli matrices acting on the EL-H and pseudo-spin subspaces, respectively.

In $\HBHZ$, $2M$ is the energy difference between the EL and H levels of the QW at the $\Gamma$ point. When the ratio $M/B$ changes sign, a band inversion occurs, what discriminates between TI and normal insulating phases. 
In \textit{asymmetric} InAs/GaSb QWs this can be achieved by applying an electric field along the growth direction of the heterostructure, modifying the sign of $\DG$ (or equivalently $M$ in the BHZ model) via the linear Stark shift effect \cite{LiuPRL08,KnezPRL11}. However, in the present inverted regime, the collapse of $\Dh$ occurs without changing the sign of $\DG$, which actually increases with the field, as seen in Fig.~\ref{fig:fig2}(a). Therefore, for ease of analysis we keep $M$ constant with respect to the field. We also anticipate that the Hamiltonian in Ref.~\cite{rothe2010} includes additional Rashba coupling terms between states $\left\{\ket{\text{EL}+},\ket{\text{EL}-}\right\}$ and between states $ \left\{\ket{\text{H}+},\ket{\text{H}}-\right\}$, which we neglect for now. Their effect will be considered later. 

To reproduce full \kp calculations with the minimal representation $H_4$, fitting is enforced at the crossing point (see Sec.~\ref{sec:lateral confinement} of the SI). In Fig.~\ref{fig:fig3}(c) we compare the exact \kp calculations with the eigenvalues of $H_{4}(k_z)$. Note that $H_4$ is accurate from the crossing point down to $\Gamma$ (much less so for the other two subbands, since interactions with higher energy states, not included in $\HBHZ$, play a major role).
Diagonalization of $H_{4}(k_z=k_0)$ shows that gap vanishing occurs as 
\begin{subequations}
\begin{align}
    \mathcal{M'}(k_{0}) &= 0 \,, \label{eq:symmetry_cond_1}  \\
    \mathcal{A} k_{0} + S_0 \left(  k_0^2 - \frac{\pi^2}{L_x^2} \right) &= 0 \label{eq:symmetry_cond_2} \, .
\end{align}
\label{eq:symmetry_conditions}
\end{subequations} 

This is confirmed in the inset of Fig.~\ref{fig:fig3}(c), where we show that the gap $\Dh$ (solid line), which is in fair agreement with the 8-band \kp calculations in Fig~\ref{fig:fig2}. The collapse at a finite wave vector $k_0$ is accompanied by the vanishing of the left-hand side of Eq.~(\ref{eq:symmetry_cond_2}) (dashed line), which means that \textit{the same-spin electron-hole kinetic interaction $\mathcal{A} k_{0}$ is exactly compensated by the opposite-spin electron-hole Rashba term proportional to $S_0$.} Figure \ref{fig:fig3}(c) also reports the spin component along $x$, showing that at the crossing point the subbands are fully spin polarized in opposite directions, forced by the $x$-directed SO field induced by $E_y$.

It is natural to ask which is the relationship between the predicted field-induced semimetallic phase at finite $k_0$ and the QSH phase occurring at $\Gamma$ in topological QWs. Using a large basis set of functions $\psi_n(x)$ to simulate increasingly large strips, we checked that $\overline{E}_+$ and $k_0$ both evolve  to zero as $L_x\rightarrow\infty $. Note that, contrary to the Dirac point of the QSH state, at finite $k_0$ the PH parity is slightly removed, not being enforced by Kramers degeneracy as at the $\Gamma$ point. Indeed, it is straightforward to show (see Sec.~\ref{sec:two-by-two hamiltonian} of the SI) that $H_{4}(k_z)$ in the degenerate subspace at $k_0$ $\left\{\ket{\psi_{1}},\ket{\psi_{2}}\right\}$ reads
\begin{equation}
   {H}_{2}(k) = E_{0} \sigma_0 - \left[2 k_0 D \sigma_0 +2 k_0 B \sigma_y  + \left( \mathcal{A} +2 S_0 k_0 \right) \sigma_x \right] k \,,
\label{eq:BHZ projected_n=1 2x2 close k0 rotated}
\end{equation}
being $k=k_z - k_0$, with the energy spectrum $E_{\pm}(k) = E_0 + \hbar v_{\pm} k $, where 
\begin{equation}
    \hbar v_{\pm} =  -2 k_0 D \pm \sqrt{4 k_0^2(S_0^2 + B^2) + 4 S_0 k_0 \mathcal{A} + \mathcal{A}^2 } \,.
\end{equation}
Hence, $|v_+| \neq |v_-|$ if $D\neq 0$, i.e. if electrons and holes have different masses. 
Using parameters in Tab.~\ref{tab:param_BHZ} of the SI, $v_+=1.326 \times 10^5$m/s and $v_-=-1.226 \times 10^5$m/s.

\begin{figure}[ht]
    \centering
    \includegraphics[width=1\columnwidth]{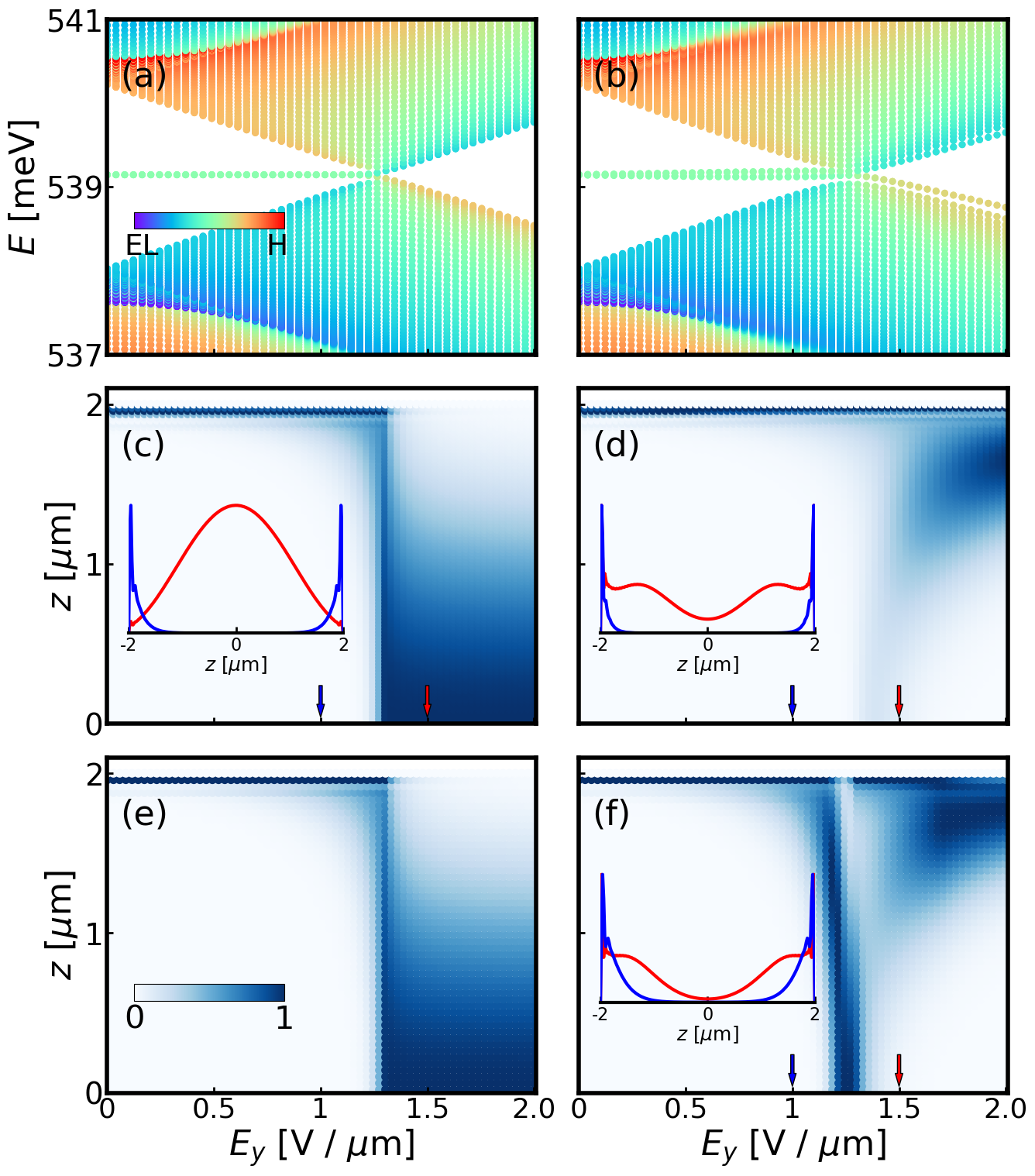}
    \caption{Energy spectra for a finite length NW with $L_z=2\,\mu\text{m}$ \textit{vs} $E_y$ (a,b) and localization of corresponding end states (c-f). Left (a,c,e): neglecting $H_R$. Right (b,d,f): including $H_R$, with $\kappa_{el}=5$ nm$^{2}$ and $\kappa_{h}=50$ nm$^{4}$, see text. In (a,b) hue represents the EL/H contribution to each state. In (c-f) hue represents the square modulus of the spinorial envelope function, normalized to its maximum. Panel (c,e) are identical due to end state degeneracy without $H_R$. Insets to panels (c,d,f): profiles at the two fields indicated by the arrows. (c) and (d) refer to the lowest energy Kramer's pair, (e) and (f) to the highest energy Kramer's pair.}
    \label{fig:fig4}
\end{figure}

The existence of gapless modes localized at the boundaries is a defining property of TIs. 1D systems which only obey TR symmetry, but do not possess PH symmetry, are not expected to support such states \cite{StanescuBook}. However, based on the BHZ Hamiltonian, mid-gap states strongly localized at the wire extremes --aka end states-- have been predicted in InAs/GaSb NWs in the inverted regime \cite{VinasPRB20}. 
As we show below, end states are supported also when the inversion symmetry is removed by a field only up to gap collapse, after which they disappear. 

To show this, we study NWs with a finite length $L_z \sim \mu \mathrm{m}$. We represent $H_4 (k_z)$ on a basis of $N_s\sim 10^2$ sine functions in the $z$ direction,
$
    \psi_n(z) = \sqrt{{2}/{L_z}} \sin\left({n \pi }/{L_z} \left(z-{L_z}/{2}\right)\right),
$
and we diagonalize the resulting $4 N_s$ Hamiltonian matrix. In Fig.~\ref{fig:fig4}(a) we show the low-energy discrete, but dense eigenvalues $vs$ $E_y$. All states are doubly degenerate due to TR symmetry and are distributed in two gapped bands, except at $\overline{E}_+$, where the gap vanishes. The states at the gap edges are strongly hybridized, as seen by the EL/H character in color code, as they arise from the discretization of states at the hybridization gap. Note that the prevalent H or EL character ($\sim$ 40\% - 60\%) at the band edges is in the inverted order for $E_y<\overline{E}_+$. In this regime two almost degenerate doublets \footnote{A small splitting, not visible in Fig.~\ref{fig:fig4}(a), is due to the slight PH asymmetry, see discussion of Eq.~\ref{eq:BHZ projected_n=1 2x2 close k0 rotated}} of states appear in the middle of the gap, with almost exact 50\%-50\% EL-H character. These states are strongly localized at the NW extremes, as seen in Fig.~\ref{fig:fig4}(c,e), and do not disperse with the field \footnote{This is actually a result of the fact that $M$ is kept constant in Eq~\ref{eq:BHZ_projected_n=1}. In reality, the band edges would move unevenly with the field with respect to the middle of the gap, see Fig.~\ref{fig:fig2}(a), and we expect the end states to slightly disperse, still being located in the middle of the gap. This would be captured by full \kp calculations for finite NWs which, however, would be prohibitive for exact diagonalization techniques.}. Accordingly, also the localization length is not affected by the field. 

As seen in Fig.~\ref{fig:fig4}(a), end states persist up to the critical field $\overline{E}_+$. For larger fields, when the gap reopens, they split in two Kramers degenerate doublets which merge with the edges of the corresponding EL or H bands, which are still strongly hybridized, but they are now in the normal order. The envelope function  (Fig.~\ref{fig:fig4})(c,e) confirms that these states are completely delocalized along the NW. Comparing Figs.~\ref{fig:fig4}(c,e) one can also note the almost perfect PH symmetry between the two bands.

We next add to $\HBHZ$ a Rashba term of odd order in $k_-$, $H_R = \left(\begin{matrix} 0 & h_R \\ h_R^\dag & 0 \end{matrix}\right)$ with 
$
 h_R = \left(\begin{matrix} -i R_0 k_- & 0 \\ 0 & i T_0 k_-^3 \end{matrix}\right)   
$
which has been left out in Eq.~(\ref{eq:BHZ_hamiltonian_2d}). Both $R_{0}$ and $T_{0}$ are linear functions of the electric field. Below we use $R_{0}/(eE_y) = 5\,\text{nm}^{2}$ and $T_0/(eE_y) = 50 \,\text{nm}^{4}$. In Fig.~\ref{fig:fig4}(b) we report the states for a the same NW as in panel (a). Since coupling terms in $h_R$ are different for $\ket{\text{EL}\pm}$  and $\ket{\text{H}\pm}$  pairs, $H_R$  removes further the PH symmetry, which in turn results in a sizable splitting between the doublets of end states at $E_y < \overline{E}_+$. At $E_y < \overline{E}_+$, after the gap collapse and reopening, end states move to, but keep slightly split from, the VB and CB states, the splitting being larger for holes, see Fig.~\ref{fig:fig4}(b). Consistently, after delocalization at the critical field, the envelope function of both pairs (see Figs.~\ref{fig:fig4})(e,f) corresponds to surface states localized at the two ends of the NW. Unlike the end states at lower fields, the localization length is field dependent and much broader. All in all, $H_R$ makes the transition less abrupt, with substantially different localization of EL and H states, as shown by the comparison of Figs.~\ref{fig:fig4}(e-f). Also note the different localization of the two doublets at $E_y\sim\overline{E}_+$.

To summarize, we have identified a transition through a semimetal state in a prototype quasi-1D system, which is a TI in 2D. Full self-consistent 8-band \kp calculations combined with the analysis of a BHZ Hamiltonian show that the collapse of $\Dh$ is induced by SO coupling, as controlled by the field, and occurs at two critical fields $\overline{E}_\pm$ between opposite spin-polarized bands at Kramers-related Dirac points. This differs from the 2D case, where the transition is driven by electrostatic effects at the $\Gamma$ point \cite{LiuPRL08,DeMedeiros2021}.
It turns out that for fields in the range $(\overline{E}_-,\overline{E}_+)$ doublets of localized, almost dispersionless end states are supported between band inverted states. However, as the field is swept across $\overline{E}_\pm$, end states first delocalize abruptly along the NW, and eventually evolve to trivial states which linearly disperse with the field, next to the VB and the CB edges. Interestingly, \textit{the transition to the trivial phase is only due to field-controlled SO coupling}, and happens without altering the sign of the inverted gap at $\Gamma$, $\Delta_\Gamma$. 

Note that no symmetry is implied in our calculations, and gap collapse occurs at a critical field regardless of structural parameters. For example, the above analysis has been performed with only one pair of inverted GaSb and InAs subbands, which requires engineering of sufficiently thin core and shell. The required parameters are accessible to current growth technologies \cite{Wang2024Vac}. However, in 
Fig.~S3 
we show that massless Dirac points may also occur when more subbands cross, which allows for more flexible realization and detection of the field-guided semimetal phase. Further flexibility might be given by ultra-strong radial electric fields obtained, for instance, by ion gating \cite{Prete2021}. The occurrence of the semimetal phase is also robust with respect to the crystallographic direction, e.g. with the field applied corner-to-corner (see Fig.~S4 in SI). 
Furthermore, since the predicted semimetal state between topologically inequivalent phases rely on the inversion asymmetry controlled by an electric field in a spatially separated electron-hole system, we expect that a similar physics can be realized in quasi-1D systems obtained from gated InAs/GaSb QWs \cite{Meyer2023}.

Finally, we note that crossing subbands, say, at $+k_0$ have opposite spin orientation [see Fig.~\ref{fig:fig3}(c)], while a degenerate same spin-propagating partner is in proximity of $-k_0$. Hence we expect scattering at the Dirac point to be  strongly suppressed \cite{McEuan99}. In a different perspective, a field-tunable Dirac point may extend the study of topological indirect excitons \cite{Du2017} to 1D systems. 

\section*{Acknowledgments}
We thank Elsa Prada, Massimo Rontani, Andrea Secchi, Pawel Woijcik and Francesco Rossella for insightful discussions. We acknowledge partial support from EU Horizon Europe research and innovation programme under the Marie Sk\l{}odowska-Curie grant agreement number 101120240. 


\newpage

\input{supporting_material}

\bibliography{bibliography}

\end{document}

%% file: supporting_material.tex



\section*{Supplementary Information to "Dirac points and end states controlled by spin-orbit coupling in inverted gap nanowires"}
This Section contains additional information on the theoretical modelling and additional \kp calculations for selected structures.




\section{\kp calculations}
\label{sec:kp calculations}

We consider an hexagonal InAs-GaSb core-shell nanowire (NW) grown along the [111] crystallographic direction, assigned to the $z$ direction, and cross-section in the $(x,y)$ plane, $\boldsymbol{r}_{\perp}=x \boldsymbol{e}_x + y \boldsymbol{e}_y$ being the in-plane position vector. To calculate its band structure we employ the 8-band \kp model~\cite{Bahder:PRB90} within the envelope function approximation~\cite{Bastard:book90}.
The NW is assumed to be translational invariant along the $z$-axis. As a consequence, the axial wave vector $k_z$ is a good quantum number, and the wave function can be written as
\begin{equation}
    \Psi(\boldsymbol{r}, k_z)=\sum_{\nu=1}^{8} e^{i k_z z} \psi_{\nu}(\boldsymbol{r}_{\perp}) \ket{\nu}
\end{equation}
where $\ket{\nu}$ are the Kane basis functions,
\begin{equation}
    \begin{split}
         \ket{1} &= \ket{\frac{1}{2},\frac{1}{2}}_{\rm EL}, \ket{2} = \ket{\frac{1}{2},-\frac{1}{2}}_{\rm EL}, \\ \ket{3} &= \ket{\frac{3}{2},\frac{3}{2}}_{\rm HH}, \ket{4} = \ket{\frac{3}{2},-\frac{3}{2}}_{\rm HH}, \\\ket{5} &= \ket{\frac{3}{2}, \frac{1}{2}}_{\rm LH}, \ket{6} = \ket{\frac{3}{2},-\frac{1}{2}}_{\rm LH}, \\ \ket{7} &=\ket{\frac{1}{2},\frac{1}{2}}_{\rm SO}, \ket{8} =\ket{\frac{1}{2},-\frac{1}{2}}_{\rm SO}  ,
     \end{split}
     \label{8BM-spin}
\end{equation}
being the basis states $\ket{\nu} = \ket{J,M_{J}}$ labelled by the Bloch total angular momentum $J$ and its component along the $z$-axis $M_J$. The subscript EL refers to the $s$-like conduction band (CB) Bloch orbitals, while the labels HH, LH and SO correspond to the $p$-like valence band (VB) Bloch orbitals, also referred to as heavy, light and split-off holes, respectively. The energy subbands $E(k_z)$ and the in-plane envelope function components $\psi_{\nu}(\boldsymbol{r}_{\perp})$ are obtained by a self-consistent numerical solution of the coupled envelope function equations, $\left( H_{\rm 8B}\right)_{\mu \nu} \psi_{\nu} = E \psi_{\mu}$, and the two-dimensional (2D) Poisson equation  $\nabla^2\phi(\boldsymbol{r}_{\perp})=-\rho(\boldsymbol{r}_{\perp})/\epsilon$, where $\phi(\boldsymbol{r}_{\perp})$ is the electrostatic potential inside the wire, 
\begin{equation}
    \rho(\boldsymbol{r}_{\perp}) = e \left( n_h(\boldsymbol{r}_{\perp}) - n_e(\boldsymbol{r}_{\perp}) \right)
\end{equation} 
is the charge density corresponding to the occupied states, being $n_e(\boldsymbol{r}_{\perp})$ and $ n_h(\boldsymbol{r}_{\perp})$ the free electron and hole carrier densities, respectively, while $\epsilon=\epsilon_r\epsilon_0$ is the dielectric constant.
In our calculations a transverse external electric field $\boldsymbol{E} = E_x \boldsymbol{e}_x + E_y \boldsymbol{e}_y$ is included by adding the term 
\begin{equation}
    H_{E} = -e \boldsymbol{E} \cdot \boldsymbol{r}_{\perp} 
    \label{eq:external electric field}
\end{equation}
to the diagonal of the 8-band Hamiltonian $H_{\rm 8B}$. For the results presented in the main text we have considered $\boldsymbol{E} = E_{y} \boldsymbol{e}_y$, thus $H_{E}=-e E_y y$ in that case.

InAs/GaSb heterostructures are broken-gap semiconductors, and the evaluation of the charge density $\rho(\boldsymbol{r}_{\perp})$ requires a special care. Here, the overlap of CBs and VBs states lead to the formation of hybridized electron-hole states in the vicinity of the chemical potential, what rules out the possibility to unequivocally occupy each state with either electrons or holes. In this work, we calculate $n_e(\boldsymbol{r}_{\perp})$ and $n_h(\boldsymbol{r}_{\perp})$ separately, using an heuristic method that we developed by extending previous work on 2D InAs/GaSb heterostructures \cite{lapushkin2004self}.

The parameters of the 8-band model we use for the simulations are reported in Table~\ref{tab:table_params}. Here, the modified Luttinger parameters are calculated as \cite{Birner2014}
\begin{equation}
\begin{split}
 \widetilde{\gamma}_1 & = \gamma_1 - \frac{E_{\rm P}}{3 \Delta_{\rm g}}\,, \\
 \widetilde{\gamma}_2 & = \gamma_2 - \frac{E_{\rm P}}{6 \Delta_{\rm g}}\,,  \\
 \widetilde{\gamma}_3 & = \gamma_3 - \frac{E_{\rm P}}{6\Delta_{\rm g}}\,,
\end{split}
\label{eq:luparam}
\end{equation}
where$\gamma_i$ are the material-dependent Luttinger parameters, $E_{\rm P}$ is the Kane energy and $\Delta_{\rm g}$ is the semiconductor bulk gap. As suggested by Foreman in Ref.~\cite{Foreman1997}, in order to avoid spurious solutions in the 8-band k$\cdot$p theory, we take a modified definition for the Kane energy parameter, i.e.,
\begin{equation}
    E_{\rm P}^{\text{mod}} = \frac{\Delta_{\rm g} (\Delta_{\rm g} + \Delta_{\rm so})}{\Delta_{\rm g} + \frac{2}{3} \Delta_{\rm so}} \left( {\frac{m_{0}}{m_e}} \right)\,,
    \label{eq:modified kane energy}
\end{equation}
where $m_0$ is the free electron mass, $m_e$ is the CB effective electron mass and $\Delta_{\rm so}$ is the split-off gap of the bulk semiconductor. Note that the modified Luttinger parameters $\widetilde{\gamma}_{i}$ are calculated from Eqs.~(\ref{eq:luparam}) with $E_{\rm P}$=$E_{\rm P}^{\text{mod}}$.

We solve the coupled envelope function equations and the Poisson equation using a real-space finite element method (FEM)~\cite{VezzosiPRB2022}. Particular care has been used in choosing the grid, which in our 2D FEM implementation consists of an arbitrary collection of triangular elements. First, we used a 2D hexagonal domain with a (non uniform) $D_{6}$ symmetry-compliant mesh of triangular elements. Using a discretization that is compliant with the symmetry of the confinement potential allows to reproduce the expected orbital degeneracies of the hexagonal domain. Second, being interested in SO effects, care has been taken in using a centro-symmetric triangular grid to avoid artificial, grid-induced contributions to Rashba splittings. A typical grid consist of 1300 elements and 700 nodes. Finally, to avoid the emergence of spurious solutions induced by the discretization \cite{spurious_vogl}, we use a mixed polynomial basis, where third-order Hermite polynomials are used for the $s$-like CB components and second-order Lagrange polynomials are used for for the $p$-like VB components \cite{Ma2014}.

Finally, we evaluate the electron and hole character of the subband states as the surface integral of the square modulus of the envelope function on the InAs and GaSb layer, respectively. In particular,
\begin{equation}
    \text{EL} = \int \text{d} \rprp \sum_{\mu} | \psi_{\mu}(\rprp, k_z) |^2 \Theta_{\text{InAs}}(\rprp) \,,
    \label{eq:electron_character}
\end{equation}
\begin{equation}
    \text{H} = \int \text{d} \rprp \sum_{\mu} | \psi_{\mu}(\rprp,k_z) |^2 \Theta_{\text{GaSb}}(\rprp)\,,
    \label{eq:hole_character}
\end{equation}
where the $\Theta_{\text{InAs}}$ and $\Theta_{\text{GaSb}}$ are the step functions 
\begin{equation}
\Theta_{\text{InAs}}(\rprp)=
    \begin{cases}
        1 \quad \text{if} \quad\rprp \in \text{InAs core}\\
        0 \quad \text{if} \quad \text{otherwise}
    \end{cases}\,,
\end{equation}

\begin{equation}
\Theta_{\text{GaSb}}(\rprp)=
    \begin{cases}
        1 \quad \text{if} \quad \rprp \in \text{GaSb shell}\\
        0 \quad \text{if} \quad \text{otherwise} 
    \end{cases}\,.
\end{equation}

\setlength{\arrayrulewidth}{.1em}
\begin{table}[b]
\begin{center}
\begin{tabular}{lccc}
\hline
\textrm{}&
\textrm{InAs}&
\textrm{}&
\textrm{GaSb}\\
\hline
$\Delta_{\rm g}$ [eV] & 0.417 &  & 0.812\\
$\delta E_{\rm c}$ [eV] & & 0.955 & \\
$\delta E_{\rm v}$ [eV] &  & 0.56 & \\
$\Delta_{\rm so}$ [eV] & 0.39 &  & 0.76 \\
$E_{\rm P}~/~E_{\rm P}^{\text{mod}}$ [eV] & 21.5~/~19.1 &  & 27~/~24.8 \\
$m_e$  & 0.026 &  & 0.039 \\
$\gamma_{1}~/~\tilde{\gamma}_{1}$  & 20~/~4.7 &  & 13.4~/~3.2 \\
$\gamma_{2}~/~\tilde{\gamma}_{2}$  & 8.5~/~0.86 &  & 4.7~/~-0.39 \\
$\gamma_{3}~/~\tilde{\gamma}_{3}$  & 9.2~/~1.6 &  & 6.0~/~0.9 \\
$\epsilon_{r}$  & 15.5 &  & 15.7 \\
$T[K]$ & & 4 & \\
\hline
\end{tabular}
\end{center}
\caption{\label{tab:table_params}%
 Material parameters used in the 8-band k$\cdot$p model at temperature 4K. The bulk semiconductor energy gap $\Delta_{\rm g}$, the CB and VB offsets $\Delta E_{\rm c}$ and $\Delta E_{\rm v}$, respectively, at the GaSb/InAs interface, the split-off energy $\Delta_{\rm so}$, the Kane energy $E_{\rm P}$, the conduction electron effective mass $m_e$ and the bare Luttinger parameters $\gamma_{i}$ are taken from Ref.~\cite{vurgaftman2001}. $E_{\rm P}^{\text{mod}}$ is the modified Kane energy Eq.~(\ref{eq:modified kane energy}) and $\tilde{\gamma}_{i}$ are the modified values Eq.~(\ref{eq:luparam}). $\epsilon_r$ refers to the relative dielectric constant of each material.}
\end{table}


\section{The BHZ Hamiltonian}
\label{sec:BHZ Hamiltonian}

\setlength{\arrayrulewidth}{.1em}
\begin{table}
\begin{center}
\begin{tabular}{ c c }
 \hline
 $\mathcal{A}$ [nm eV] & 31.8 $\times$ $10^{-3}$ \\ 
 $B$ [nm$^2$ eV] & -1.052 \\  
 $D$ [nm$^2$ eV] & -0.0436 \\
 $M$ [eV] & -38 $\times$ $10^{-3}$ \\
 $C$ [eV] & 537.5 $\times$ $10^{-3}$ \\
 $\kappa$ [nm$^3$] & 27.016 \\
 $L_x$ [nm] & 16.5 \\
\hline
\end{tabular}
\end{center}
\caption{Parameters of the BHZ model Hamiltonian Eq.~(\ref{eq:BHZ_projected_n=1}). The parameter $\mathcal{A}$ regulates the strength of the electron-hole kinetic interaction. $B$ is proportional to the mean between the effective masses $m_{e/h}$, while $D$ is proportional to the difference between the two. $M$ is the value of the inverted-gap. $C$ is a constant energy shift. $\kappa$ regulates the strength of the Rashba electron-hole, opposite-spin interaction. $L_x$ is the \textit{short} strip length used to model the energy states of the NW.}
\label{tab:param_BHZ}
\end{table}
The Bernevig-Hughes-Zhang (BHZ) Hamiltonian $\HBHZ$ in Eq.~\ref{eq:BHZ_hamiltonian_2d} has been derived in Ref.~\cite{rothe2010} to describe the low-energy states of a HgTe/CdTe quantum well (QW) in the presence of an external electric field in the growth $z$-direction. This model, which neglects Rashba terms due to structural inversion symmetry, is suitable to study the low energy physics of \textit{symmetric} InAs/GaSb/InAs QWs as well. Note that our NWs, taken along a diagonal, are similar to a symmetric QW and in the absence of a transverse electric field, they lack structural asymmetry-induced Rashba terms.  

In $\HBHZ$ the parameter $C$ sets the zero of energy. The parameter $\mathcal{A}$ regulates the strength of the kinetic electron-hole interaction. The parameters $B$ and $D$ regulate the in-plane dispersion of the energy bands and can be related to the effective masses $m_e$ and $m_h$ of electrons and holes, respectively, by the relation $D \pm B =\mp \hbar^2/2 m_{e/h}$, or
\begin{equation}
D = - \frac{\hbar^2}{2} \left( \frac{1}{m_e} - \frac{1}{m_h}\right)\,, \label{eq:D parameter BHZ}
\end{equation}
\begin{equation}
B = - \frac{\hbar^2}{2} \left( \frac{1}{m_e} + \frac{1}{m_h}\right)\,,
\label{eq:B parameter BHZ}
\end{equation}
with $m_e>0$ and $m_h>0$. Note that $D$ is proportional to the difference between the effective masses, while $B$ is proportional to their average. The rest of the parameters of the Hamiltonian have been described in the main text. 

\subsection{Lateral confinement fitting procedure}
\label{sec:lateral confinement}

To model the low energy physics of the InAs-GaSb core-shell NW close to the critical fields $\overline{E}_{\pm}$, we map our quasi-1D system to a laterally confined InAs/GaSb QW as in Fig.~\ref{fig:fig3}(b). To do that we start from the BHZ Hamiltonian $\HBHZ$ in Eq.~\ref{eq:BHZ_hamiltonian_2d} and perform a relabeling of the coordinate reference frame, i.e. $(k_x,k_y)\rightarrow (k_z, k_x)$. Then, we add a lateral confinement in the $x$-direction and let the system to be translational invariant along the $z$-axis. In this case, the axial wave vector $k_z$ is a good quantum number, while $k_x$ has to be replaced with the corresponding differential operator $-i \pdv{}{x}$. Taking the length of the InAs-GaSb effective bi-layer in the confined direction $x$ to be $L_x$, we can project $\HBHZ$ onto the real-space basis
\begin{equation}
    \psi_n(x) =\sqrt{\frac{2}{L_x}} \sin\left(\frac{n \pi }{L_x} \left(x-\frac{L_x}{2}\right)\right)\,,
    \label{eq:real space basis}
\end{equation}
satisfying hard-wall boundary conditions at $\pm L_x/2$. As we aim at describing the low-energy physics of the first CB and VB energy doublets of the nanowire [see \ref{fig:fig2}(c-e)], we can consider only $n=1$ in Eq.~\ref{eq:real space basis} and, using 
\begin{equation}
\begin{split}
    \int_{-L_x /2}^{L_x/2} \, \text{d}x \, \psi_{1}(x) \left(-i \hbar \pdv{}{x}\right)^2 \psi_{1}(x) &= \left(\frac{\pi}{L_x}\right)^2 \,, \\
    \int_{-L_x /2}^{L_x/2}  \, \text{d}x \, \psi_{1}(x) \left(-i \hbar \pdv{}{x}\right) \psi_{1}(x) &= 0
\end{split}
\label{eq:matel basis}
\end{equation}
we obtain the $4 \times 4$ Hamiltonian $H_{4}(k_z)$ in Eq.~\ref{eq:BHZ_projected_n=1}.

To reproduce the full \kp calculations at the Dirac points with such a minimal basis representation, namely including only the four lowest energy subbands, as in Fig.~\ref{fig:fig3}(c), we fitted the parameters in $H_{4}(k_z)$ by the following procedure. We extract the values of gap closing wave vector, $k_0$, and energy, $E_0$, as well as the critical field $\overline{E}_+$ from the results of the 8-band model, given by the grey dashed lines in Fig.~\ref{fig:fig3}(c). Furthermore, we assume a fixed value of $\mathcal{A}$ (see Tab.~\ref{tab:param_BHZ}), close to the value employed in Ref.~\cite{Vinas2020}, and consider a value of $L_x$ consistent with the lateral dimension of the nanowire. It is easy to diagonalize the Hamiltonian $H_{4}(k_z)$ and find its eigenvalues
\begin{equation}
\begin{split}
    E_{1,2} &= \varepsilon'(k_z) \pm \sqrt{\left(\mathcal{A} k_z + S_0 \left(k_z^2 - \frac{\pi^2}{L_x^2}\right)\right)^2 + \mathcal{M}'(k_z)} \, , \\
    E_{3,4} &= \varepsilon'(k_z) \pm \sqrt{\left(\mathcal{A} k_z - S_0 \left(k_z^2 - \frac{\pi^2}{L_x^2}\right)\right)^2 + \mathcal{M}'(k_z)} \,
\end{split}
\end{equation}
Since $B>0$, $S_0>0$ and $k_0^2 < (\pi/L_x)^2$, the gap closes for $E_1=E_2=E_0$ when the expression under square root is zero, which implies
\begin{equation}
    \mathcal{M}'(k_0)=0 \, ,
    \label{eq:closing condition 1}
\end{equation}
that identifies the wave vector at the crossing points $k_0$,
\begin{equation}
    k_{0} = \sqrt{\frac{M}{B} - \frac{\pi^2}{L_x^2}} \,,
    \label{eq:k0 crossing}
\end{equation} 
and 
\begin{equation}
    \mathcal{A} k_{0} + S_0 \left(  k_0^2 - \frac{\pi^2}{L_x^2} \right) = 0 \,,
    \label{eq:closing condition 2}
\end{equation}
while the degenerate eigenvalue $E_0$ is given by
\begin{equation}
    E_0 = \varepsilon'(k_{0}) = C - D \left[ (\pi/L_x)^2 + k_0^2\right] = C - \frac{D}{B} M \,.
    \label{eq:degenerate eigenvalue}
\end{equation}
From Eq.~(\ref{eq:closing condition 2}) we can obtain the value of $\kappa$, being $S_{0}=\kappa\times e {E}_{y}$,
\begin{equation}
    \kappa = \frac{\mathcal{A}k_0}{e \Bar{E}_{y}\left(k_0^2 - \frac{\pi^2}{L_x^2}\right)} \,.
\end{equation}
Then, using Eq.~(\ref{eq:k0 crossing}) and Eq.~(\ref{eq:degenerate eigenvalue}) we eliminate the parameters $B$ and $D$ in favour of $M$ and $C$, that now represent the only two free fitting  parameters. In Tab.~\ref{tab:param_BHZ} we report the actual parameters used to produce Fig.~\ref{fig:fig3}(c). Note that the degeneracy occurs also at the Kramers related wave vector $-k_0$, where instead $E_3=E_4=E_0$, consistently with the condition in Eq.~\ref{eq:closing condition 2}.

\subsection{Two-by-two Hamiltonian at the crossing point}
\label{sec:two-by-two hamiltonian}

It is easy to show that when Eqs.~\eqref{eq:closing condition 1} and \eqref{eq:closing condition 2} are satisfied, the degenerate subspace corresponding of $H_4(k_z)$ at $k_0$ is spanned by
\begin{equation}
\begin{split}
        \ket{\psi_{1}} &= \frac{1}{\sqrt{2}} (1,0,1,0)^T\,, \\ 
    \ket{\psi_{2}} &= \frac{1}{\sqrt{2}} (0,-1,0,1)^T\,.
    \end{split}
\end{equation}
Hence, we obtained the $2 \times 2$ Hamiltonian in Eq.~\eqref{eq:BHZ projected_n=1 2x2 close k0 rotated} as follows. First, we replaced $k_z$ in favour of $k=k_z - k_0$ in $H_4$ and then projected the latter onto the degenerate subspace $\left\{ \ket{\psi_1} ,\ket{\psi_2} \right\}$ obtaining 
\begin{equation}
    H_2(k) = E_{0} \sigma_0 - \left[2 k_0 D \sigma_0 +2 k_0 B \sigma_z  + \left( \mathcal{A} +2 S_0 k_0 \right) \sigma_x \right] k \,,
\end{equation}
After that, we rotated the coordinate system using the operator $U=\exp \left({i \frac{\varphi}{2}\sigma_x}\right)$ with $\varphi=\pi/2$ and obtained the rotated Hamiltonian $\widetilde{H}_2 = U H_2 U^{\dag}$, which is the one that appears in the main text as $H_2$.

\section{Inclusion of Rashba terms}
\label{sec:rashba terms}

As stated in the main text, the Rashba term of odd order in $k_-$ reads
\begin{equation}
    H_R = 
    \begin{pmatrix} 
    0 & h_R \\ 
    h_R^{\dag} & 0 
    \end{pmatrix} \,,
\end{equation}
with 
\begin{equation}
h_R = \left(\begin{matrix} -i R_0 k_- & 0 \\ 0 & i T_0 k_-^3 \end{matrix}\right) \,.
\end{equation}
The corresponding term in the laterally confined QW model is obtained as in Sec.~\ref{sec:lateral confinement}. Using Eq.~(\ref{eq:matel basis}) and
\begin{equation}
    \int_{-L_x /2}^{L_x/2} \, \text{d}x \, \psi_{1}(x) \left(-i \hbar \pdv{}{x}\right)^3 \psi_{1}(x) = 0 \,,
\end{equation}
we obtain
\begin{equation}
\begin{split}
    H_{R,4} &= \frac{R_0 k_z}{2} \sigma_y \left( \tau_0 + \tau_z \right) - T_0 \frac{k_z^3 -3 k_z \pi^2/L_x^2}{2} \sigma_y \left( \tau_0 - \tau_z \right) \\
    &= \mathcal{R}_{+} \sigma_y \tau_0 + \mathcal{R}_{-} \sigma_y \tau_z
\end{split}
\label{eq:rashba effective}
\end{equation}
where
\begin{align}
    \mathcal{R}_{+} &= \frac{k_z (R_0 + 3 \pi^2 T_0 /L_x^2) - T_0 k_z^3}{2}\,, \\
    \mathcal{R}_{-} &= \frac{k_z (R_0 - 3 \pi^2 T_0/L_x^2) + T_0 k_z^3}{2}\,.
\end{align}

As discussed in the main text, 1D systems which obey TR symmetry, but do not possess PH symmetry, have no topological class. For example, in the effective BHZ Hamiltonian in Eq.~(\ref{eq:BHZ_projected_n=1}), PH symmetry is broken by the term proportional to the identity operator, namely $D \mathds{1}_4$, which arises from the difference between the electron and hole effective masses. The additional Rashba term $H_{R,4}$ in Eq.~(\ref{eq:rashba effective})  discussed here also breaks the PH symmetry. However, it is easy to show that PH can be recovered in the model Hamiltonian if $D=0$ (which corresponds to $m_{el}=m_h$) and $\mathcal{R}_{+}=0$ \textit{or} $\mathcal{R}_{-}=0$.\footnote{$\mathcal{R}_{+}=\mathcal{R}_{-}=0$ in not possible, being $\mathcal{R}_{+}$ and $\mathcal{R}_{-}$ the sum and the difference of two terms}
In fact, under this assumption, it is possible to find a unitary transformation $S$ such that
\begin{equation}
    S H_{4} S^{\dag} = \begin{pmatrix}
        0 & h_c \\
        h_c^{\dag} & 0
    \end{pmatrix}
\end{equation}
or, equivalently, to show that the Hamiltonian possesses chiral-symmetry, being $\mathcal{C}$ the corresponding chiral-symmetry operator. Being also TR symmetry present with $\mathcal{T}=\sigma_y \tau_0$, the Hamiltonian is also PH symmetric, with the corresponding operator given by a combination of the latter, i.e. $\mathcal{P} = \mathcal{C} \mathcal{T} $. If $\mathcal{R}_{+}=0$ we have
\begin{equation}
    S = \frac{1}{\sqrt{2}}\begin{pmatrix}
        -i & 0 & 0 & 1 \\
        0 & 1 & i & 0 \\
        1 & 0 & 0 & -i \\
        0 & 1 & -i & 0
    \end{pmatrix} \,,
\end{equation}
whose columns are the eigenvectors of the chiral-symmetry operator $\mathcal{C}_1 = \sigma_y \tau_x$, with $\mathcal{P}^2=-1$. For $\mathcal{R}_{-}=0$ instead we have
\begin{equation}
    S= \frac{1}{\sqrt{2}}\begin{pmatrix}
        -i & 1 & 0 & 0 \\
        0 & 0 & i & 1 \\
        1 & -i & 0 & 0 \\
        0 & 0 & 1 & i
    \end{pmatrix} \,,
\end{equation}
whose columns are the eigenvectors of a second choice for the chiral-symmetry operator~\cite{Matveeva:PRB23} $\mathcal{C}_2 = \sigma_y \tau_x$, with $\mathcal{P}^2=1$. If $D\neq0$, PH symmetry is broken due to the term proportional to the identity matrix $\sigma_0 \tau_0$ in $H_4$. Furthermore, if $\mathcal{R}_+ \neq 0$ and/or $\mathcal{R}_- \neq 0$, the PH symmetry is also removed, either by the term $\sigma_y \tau_0$, if $\mathcal{C} = \mathcal{C}_1$, or by the term $\sigma_y \tau_z$, if $\mathcal{C} = \mathcal{C}_2$.


\section{The role of self-consistent effects}
\label{sec:role of SCF}
\label{sec:self-consistent effects}

\begin{figure}[ht]
    \centering
    \includegraphics[width=0.4\columnwidth]{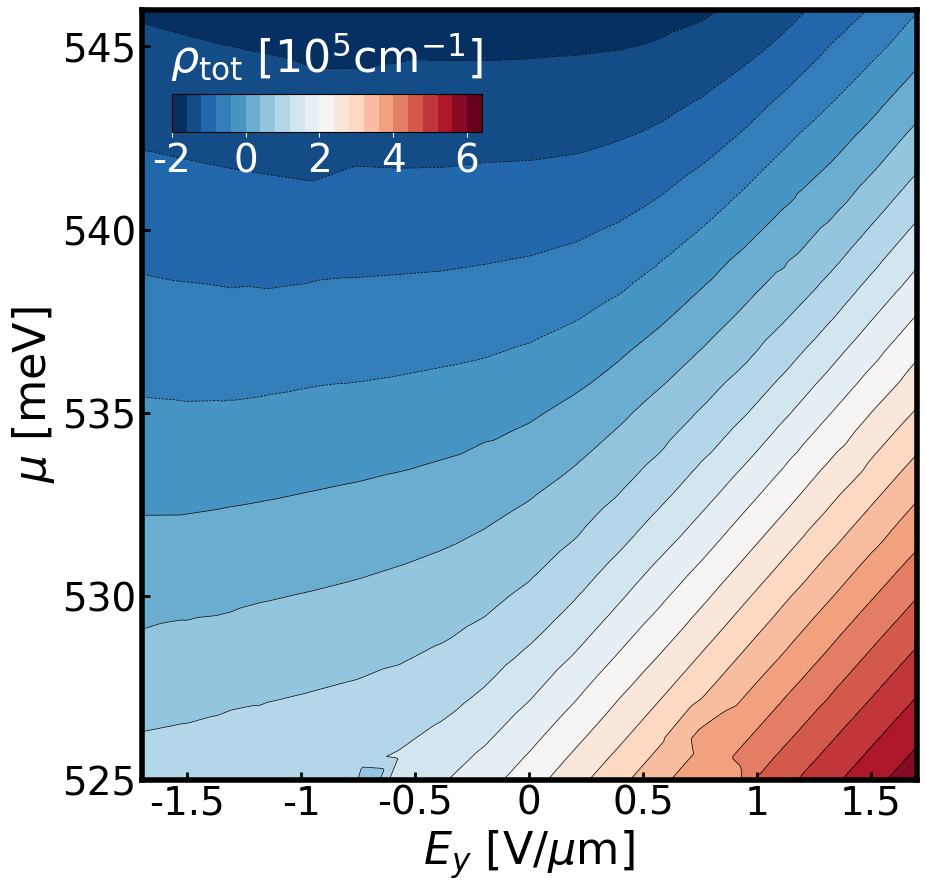}
    \caption{Total charge density $\rho_{\rm tot}$ [see Eq.~\eqref{eq:total charge density}] of the InAs/GaSb core-shell NW with core radius $R_c=7$ nm and shell width $w=4.88$ nm as a function the chemical potential $\mu$ and the transverse electric field $E_{y}$.}
    \label{fig:total charge density}
\end{figure}

As explained in Sec.~\ref{sec:kp calculations}, the subband energy levels are calculated by self-consistently solving the 8-band \kp model envelope function equations and the Poisson equation. The latter is solved for the built-in electrostatic potential $\phi(\boldsymbol{r}_{\perp})$, which is non-zero due to the presence of a charge density $\rho(\boldsymbol{r}_{\perp})$ of free electron and hole carriers in the NW. The total charge is computed from 
\begin{equation}
    \rho_{tot} = \int_{\Omega} \text{d}\boldsymbol{r}_{\perp} \left( n_h(\boldsymbol{r}_{\perp}) - n_e(\boldsymbol{r}_{\perp}) \right) \,,
    \label{eq:total charge density}
\end{equation}
where $\Omega$ is the 2D section of the nanowire. In Fig.~\ref{fig:total charge density} we report $\rho_{\rm tot}$ as a function of the chemical potential $\mu$ and $E_y$, covering both negatively and positively charged regimes, separated by a charge neutrality point which shifts linearly with positive electric fields.

\begin{figure}[ht]
    \centering
    \includegraphics[width=0.4\columnwidth]{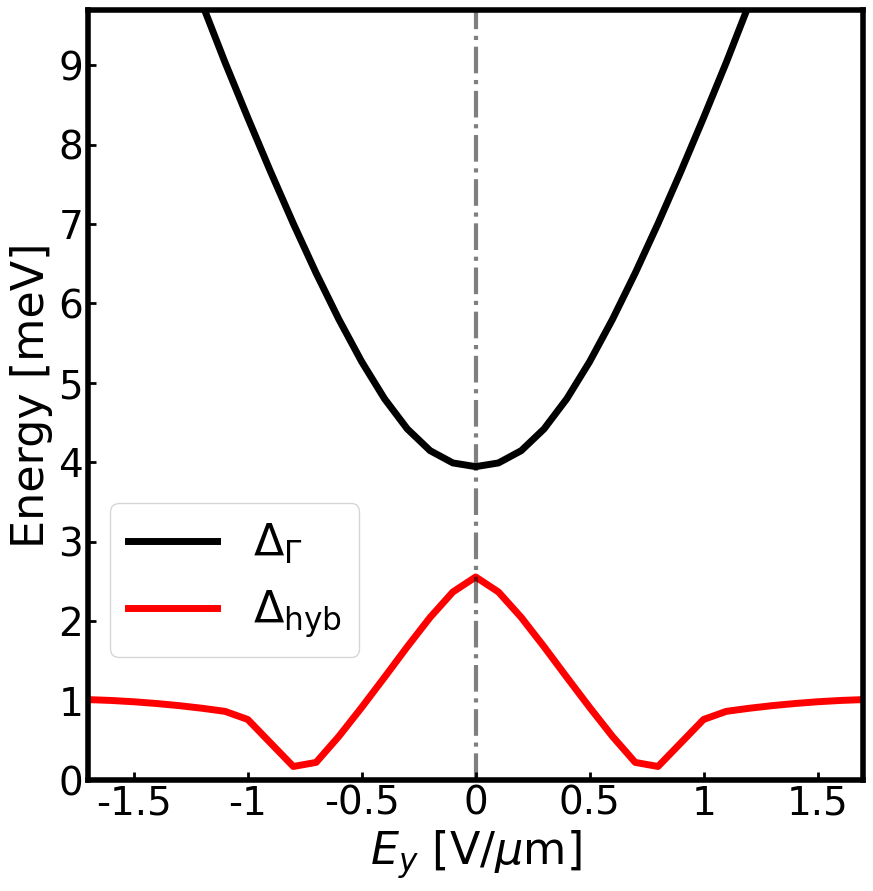}
    \caption{Inverted gap $\Delta_{\Gamma}$ (black) and hybridization gap $\Dh$ (red) as a function of the transverse electric field $E_{y}$ without including self-consistent effects for the same NW as in Fig.~\ref{fig:fig2}}
    \label{fig:no self cons}
\end{figure}

In Fig.~\ref{fig:no self cons} we show the values of the inverted gap $\DG$ and the hybridization gap $\Dh$ as a function of the transverse electric field $\boldsymbol{E}=E_y \boldsymbol{e}_y$ (the analogous of Fig.~\ref{fig:fig2}(a)) without including the self-consistent field. In this case, the external potential is generated solely by the transverse electric field, while the built-in electric field originated by the non-zero charge density inside the NW (which is present also at zero field, due to charge transfer between the layers) is neglected. As a result, the evolution of $\DG$ and $\Dh$ is symmetric with respect to $E_y=0$. Close to $E_y=0$, $\DG$ increases quadratically with $E_y$, while at strong electric fields it increases linearly with $E_y$. 
The hybridization gap $\Dh$ first decreases with $E_y$, and importantly, vanishes at two symmetric values of the electric field $\overline{E}_\pm = \pm E_0$, analogously to the full calculations in Fig.~\ref{fig:fig2}(a). For $|E_y|>E_0$, $\Dh$ increases again and saturates at large electric fields. We conclude that self-consistent effects are not qualitatively essential to the physics of the predicted semimetal phase, consistently with the physical SO origin of the transition; however they play a significant role in quantitative terms, as the values of $\DG$, $\Dh$ and $\overline{E}_\pm$ are actually influenced by the built-in electric field.

\section{Tilted electric field}
\label{sec:tilted field}

In this section we consider an electric field which is directed corner-to-corner with respect to the NW hexagonal symmetry, i.e. forming an angle $\theta = \pi/6$ with respect to the positive direction of the $y$-axis, i.e. $\boldsymbol{E} = E_{{\pi}/{6}} \left( - \frac{1}{2} \mathbf{e}_x + \frac{\sqrt{3}}{2} \mathbf{e}_y \right)$, with intensity $E_{{\pi}/{6}}$.

\begin{figure}[ht]
    \centering
    \includegraphics[width=0.5\columnwidth]{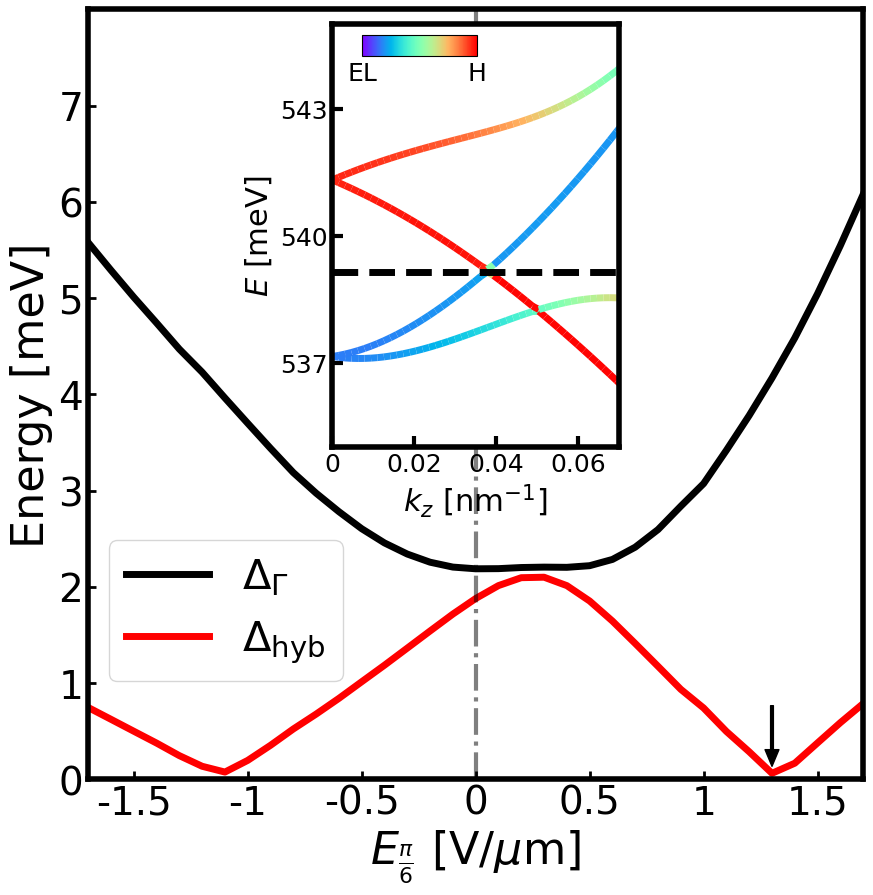}
    \caption{Inverted gap $\Delta_{\Gamma}$ (black) and hybridization gap $\Dh$ (red) as a function of the tilted electric field intensity $E_{{\pi}/{6}}$ for the same NW discussed in the main text. Inset: subband dispersion at the critical electric field indicated by the black arrow in the main panel.}
    \label{fig:tilted}
\end{figure}

In Fig.~\ref{fig:tilted} we plot $\DG$ and $\Dh$ as a function of $E_{{\pi}/{6}}$ with the same chemical potential $\mu$ used in Fig.~\ref{fig:fig2}(a,d-e). Importantly, we do not observe any significant qualitative difference with respect to the case with a vertical (i.e. facet-to-facet) electric field discussed in the main text. For completeness, the inset of Fig.~\ref{fig:tilted} shows the band structure for the positive critical field $\Bar{E}_{{\pi}/{6}_{+}}$, indicated by the black arrow. 

We note that the corner-to-corner field configuration has a larger symmetry with respect to the facet-to-face configuration. Since the gap vanishes in both cases, it also occurs at any intermediate direction of the field.

\section{More than two inverted subbands}
\label{sec:more inverted subbands}

In the main text we have considered the semimetal phase with only one pair of inverted electron and hole subbands at zero field, which requires fine tuning of the structural parameters. Therefore, from a practical point of view it is interesting to establish whether the field-induced semimetal phase arise even with slightly different parameters, causing more than two subbands being inverted.

\begin{figure}
    \includegraphics[width=0.5\columnwidth]{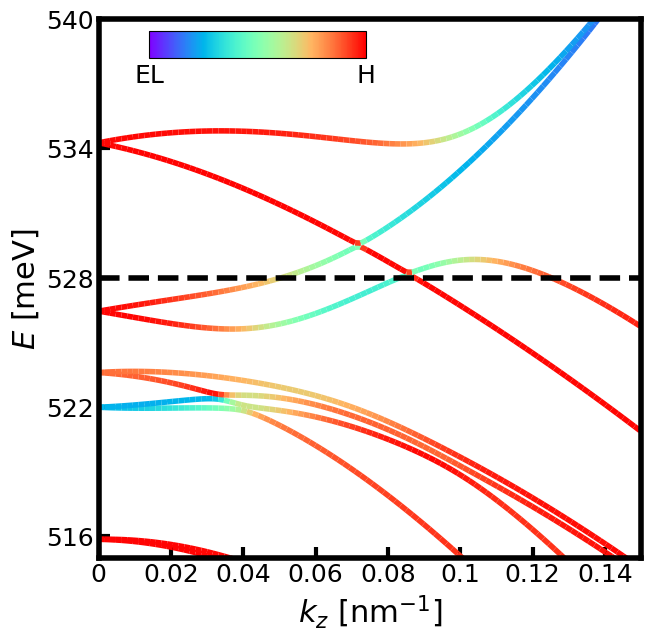}
    \caption{Subband dispersion for a NW with core radius $R_c=7.5 \,\text{nm}$ and shell width $w=4.88\,\text{nm}$ for $E_y=1$ V/$\mu$. Despite multiple valence subbands inverted with respect to a single conduction subband, a semimetal phase occurs within an otherwise overall gap, similarly to the case discussed in the main text.}
    \label{fig:more_inverted_subbands}
\end{figure}

To answer this question, we simulated a NW with $R_c=7.5$ nm and $w=4.88$ nm. For these parameters, at $E_y=0$ one electron subband invert with two hole subbands. In Fig.~\ref{fig:more_inverted_subbands} we show the energy band structure of the nanowire at $\mu=528$ meV and $E_y=1$ V/$\mu$m. As seen at the $\Gamma$ point, the lowest electron subband falls below the three highest hole subbands. At finite $k_z$ the electron subband hybridizes with the three hole subbands. Several crossings occur between the spin-split subbands, which however fall in the subband continuum. However, a hybridization gap $\Dh$ forms with the highest hole subband. Analogously to the case of only two inverted electron and hole subbands treated in the main text, the $\Dh$ collapses at a critical field in an otherwise fully gapped region.

